\documentclass[12pt,a4paper]{article}
\usepackage{amsmath}
\begin{document}

\begin{flushright}
DPNU-99-33
\end{flushright}

\vspace{30pt}

\begin{center}
{\Huge \scshape Hopf Map \medskip}

{\Huge \scshape and \medskip}

{\Huge \scshape Quantization on Sphere }{\LARGE \vspace{30pt}}

Hitoshi IKEMORI\footnote{ikemori@biwako.shiga-u.ac.jp} \\[0pt]\textit{Faculty
of Economics, Shiga University,}

\textit{Hikone, Shiga 522-8522, Japan}

\medskip

Shinsaku KITAKADO\footnote{kitakado@eken.phys.nagoya-u.ac.jp} \\[0pt]%
\textit{Department of Physics, Nagoya University,}

\textit{Nagoya 464-8602, Japan}

\medskip

Hideharu OTSU\footnote{otsu@vega.aichi-u.ac.jp} \\[0pt]\textit{Faculty of
Economics, Aichi University, }

\textit{Toyohashi, Aichi 441-8522, Japan}

\medskip

Toshiro SATO\footnote{tsato@matsusaka-u.ac.jp} \\[0pt]\textit{Faculty of
Political Science and Economics,}\\[0pt]\textit{Matsusaka University,
Matsusaka, Mie 515-8511, Japan\bigskip}

\vspace{30pt} {\Large \scshape Abstract}
\end{center}

Quantization of a system constrained to move on a sphere is considered by
taking a square root of the ``on sphere condition''. We arrive at the fibre
bundle structure of the Hopf map in the cases of $S^{2}\,$and $S^{4}$. This
leads to more geometrical understanding of monopole and instanton gauge
structures that emerge in the course of quantization.

\newpage

\section{Quantum constraints to the sphere \`{a} la Dirac}

Quantum mechanics, where the sphere $S^{n}$ embedded in $R^{n+1}$ is
considered as a configuration space, has been studied in \cite{Ohnuki:1993cb}
\cite{Ohnuki:1992uv}\cite{Ohnuki:1998xx}\cite{Ohnuki:1998yy} and the gauge
fields were seen to emerge at the quantum level, which in turn specify the
inequivalent quantizations that are possible on the sphere. The authors of
these papers, generalizing the conventional canonical commutation relations,
set up as a ``fundamental algebra'', the Lie algebra of the Euclidean group
$E(n+1)$ in $(n+1)$ dimensional space, which is given by the semidirect
product of $SO(n+1)$ and $R^{n+1}$. Then they obtain the representation of the
group using the Wigner's technique which allows them to construct the
representation of $E(n+1)$ in terms of the irreducible representation of the
`little group' $SO(n)$ -- the isometry group of $SO(n+1)$ -- acting on $S^{n}
$. Finally they show that a particle on $S^{n}$ couples to a gauge potential
covariantly through the generator of the Wigner rotation\ and that it is
related to the `induced' gauge field of the (generally) nonabelian `monopole'
located at the center of $S^{n}$.

The induced gauge fields were then shown \cite{McMullan:1995wz} to be nothing
but the $H$-connections, $i.e.$ the gauge fields that emerge when we consider
quantum mechanics on the coset space $G/H$, which were thoroughly studied in
\cite{Landsman:1991zb}. These authors consider the system of a `free particle'
on $G/H$, which in the case of sphere is $SO(n+1)/SO(n)$ , and working within
the framework of Mackey's quantization scheme \cite{Mackey:1968} have shown
that the Hamiltonian on G/H involves the induced $H$-connection.

In this paper we reconsider this problem, $i.e.$ quantum mechanics on $S^{2}$
and $S^{4}$, from somewhat different point of view. Namely we consider , \`{a}
la Dirac, a square root of the ``on sphere constraint''. The results are not
different, of course, from those obtained in the above mentioned approaches,
however, we hope to gain a deeper insight into the problem.

\section{Quantum Mechanics on $S^{2}$}

Let us start with $S^{2}$. As our sphere $S^{2}$ is embedded in the space
$R^{3}$, it is defined by the constraint
\begin{equation}
\vec{x}^{2}=a^{2}\label{sphere}%
\end{equation}
among the coordinates $\vec{x}=\left(  x_{1},x_{2},x_{3}\right)  $ in the
$R^{3}$. Naive definition of the quantum mechanics on $S^{2}$ is simply to
restrict these variables to satisfy the constraint $\vec{x}^{2}=a^{2}$ and to
require the momentum operators to be compatible with the constraint in their
Hamiltonian and the wave function. Which means that in the coordinate
representation
\begin{equation}
i\frac{\partial}{\partial t}\left[  \Psi\left(  x\right)  \right]  _{S^{2}%
}=\left[  \hat{H}\Psi\left(  x\right)  \right]  _{S^{2}}\text{ ,}%
\end{equation}
where $\left[  \;\right]  _{S^{2}}$ is the restriction of the variables on the
constraint surface $S^{2}:\vec{x}^{2}=a^{2}$. This means that for a free
particle
\begin{align}
\left[  \hat{H}\Psi\left(  x\right)  \right]  _{S^{2}} &  =\left[  -\frac
{1}{2m}\frac{\partial^{2}\;}{\partial x_{i}\partial x^{i}}\Psi\left(
x\right)  \right]  _{S^{2}}\\
&  =-\frac{1}{2m}g^{ab}\nabla_{a}\partial_{b}\left[  \Psi\left(  x\right)
\right]  _{S^{2}}\text{ ,}\nonumber
\end{align}
where $\partial_{a}$ and $\nabla_{a}$ denote the derivative and the covariant
derivative with respect to the coordinate $q^{a}$ and the metric $g_{ab}$ on
$S^{2}$ . As a result we have the Hamiltonian
\begin{equation}
\mathcal{H}=-\frac{1}{2m}g^{ab}\nabla_{a}\partial_{b}\text{ ,}%
\end{equation}
which acts on the wave function $\psi\left(  q\right)  =\left[  \Psi\left(
x\right)  \right]  _{S^{2}}$ .

In this paper we propose a new definition of the quantum mechanics on $S^{2}$
replacing (\ref{sphere}) by the ``quantum constraint'' on the wave function
\begin{equation}
\left(  \vec{x}\cdot\vec{\sigma}-a\right)  \Phi\left(  \vec{x}\right)
=0\text{ .}\label{quantum sphere}%
\end{equation}
This leads to
\begin{align}
\left(  \vec{x}\cdot\vec{\sigma}+a\right)  \left(  \vec{x}\cdot\vec{\sigma
}-a\right)  \Phi\left(  \vec{x}\right)   &  =0\\
\left(  \left(  \vec{x}\cdot\vec{\sigma}\right)  ^{2}-a^{2}\right)
\Phi\left(  \vec{x}\right)   &  =0\nonumber\\
\left(  \vec{x}^{2}-a^{2}\right)  \Phi\left(  \vec{x}\right)   &  =0\text{
,}\nonumber
\end{align}
thus the constraint $\;\vec{x}^{2}-a^{2}=0$ follows from the condition
(\ref{quantum sphere}). Here $\sigma$'s are the Pauli matrices. Defining
\begin{equation}
\Delta\equiv\left(  \vec{x}\cdot\vec{\sigma}-a\right)  \text{ ,}%
\end{equation}
eq.(\ref{quantum sphere}) is rewritten as
\begin{equation}
\Delta\Phi=\left(
\begin{array}
[c]{cc}%
x_{3}-a & x_{1}-ix_{2}\\
x_{1}+ix_{2} & -x_{3}-a
\end{array}
\right)  \Phi=0\text{ .}\label{quantum constraint2}%
\end{equation}
In order for this equation to have a non-trivial solution it has to be
degenerate and the determinant should vanish
\begin{equation}
\det\Delta=-\left(  \vec{x}^{2}-a^{2}\right)  =0\text{ .}%
\end{equation}
We define $v$ by
\begin{align}
\Delta v &  =0\text{ ,}\\
v^{\dag}v &  =1\text{ ,}\nonumber
\end{align}
where $v$ is $2\times1$ matrix or eigenvector of $\Delta$ whose eigenvalue is
$0$.

The explicit form of $v$ can be written as
\begin{equation}
v=\frac{1}{\sqrt{2a\left(  a+x_{3}\right)  }}\left(
\begin{array}
[c]{c}%
a+x_{3}\\
x_{1}+ix_{2}%
\end{array}
\right)  \text{ .}\label{v2}%
\end{equation}
The general solution of eq.(\ref{quantum constraint2}) can be written as
\begin{equation}
\Phi=v\phi\text{ ,}%
\end{equation}
with $\phi$ an arbitrary complex function on $S^{2}$. Thus the solution to the
constraint is the space projected by $P$%
\begin{equation}
P\Phi=\Phi\text{ ,}%
\end{equation}
where the projection operator $P$ to the space spanned by $v$ is defined as
\begin{align}
P &  \equiv vv^{\dag}\text{ ,}\\
P^{2} &  =vv^{\dag}vv^{\dag}=P\text{ ,}\nonumber\\
Pv &  =v\text{ .}\nonumber
\end{align}
Although $\Phi$ lives in this projected space, its derivative $\partial\Phi$
does not necessarily live in this space. Then our interest is concerned with
the projected derivative $P\partial\Phi\,$\textit{i.e.}
\begin{align}
P\partial\Phi & =vv^{\dag}\partial\left(  v\phi\right) \\
& =vv^{\dag}\left(  v\partial\phi+\partial v\phi\right) \nonumber\\
& =vD\phi\text{ ,}\nonumber
\end{align}
where
\begin{align}
D &  \equiv\partial+A\;\;\text{\ ,}\\
A &  \equiv v^{\dag}\partial v\;\text{\ .}\nonumber
\end{align}
It is noticed here that the gauge connection is induced as a result of this
projection. It is also obvious that for any polynomial $F\left(
\lambda\right)  $ of $\lambda$%
\begin{equation}
F\left(  P\partial\right)  \Phi=vF\left(  D\right)  \phi\text{ .}%
\end{equation}
We define the quantum mechanics on $S^{2}$ by this projection
\begin{equation}
i\frac{\partial}{\partial t}\left[  \Phi\left(  x\right)  \right]  _{S^{2}%
}=\left[  \hat{H}\left(  x,-iP\frac{\partial}{\partial x}\right)  \Phi\left(
x\right)  \right]  _{S^{2}}\text{ .}%
\end{equation}
As a result we have the Hamiltonian
\begin{equation}
\mathcal{H}=-\frac{1}{2m}g^{ab}\left(  \nabla_{a}+A_{a}\right)  \left(
\partial_{b}+A_{b}\right)  \text{ ,}%
\end{equation}
which acts on the wave function $\phi\left(  q\right)  =\left[  \Phi\left(
x\right)  \right]  _{S^{2}}$ .

Here $A$ can be written as
\begin{equation}
A\equiv v^{\dag}dv=\frac{i}{2a\left(  a+x_{3}\right)  }\left(  x_{1}%
dx_{2}-x_{2}dx_{1}\right)  \text{ ,}%
\end{equation}
and this is the induced magnetic monopole gauge potential obtained in
\cite{Ohnuki:1993cb}\cite{McMullan:1995wz}.

\section{Quantum Mechanics on $S^{4}$}

Let us turn our discussion to quantum mechanics on $S^{4}$, which is a rather
straightforward generalization of the arguments given above.

The coordinates are
\begin{equation}
x_{M}=\left(  x_{0},x_{1},x_{2},x_{3},x_{5}\right)  \text{ ,}%
\end{equation}
which are restricted by
\begin{equation}
x_{M}x^{M}-a^{2}=0\text{ .}%
\end{equation}
We choose
\begin{align}
\gamma^{M} &  =\left(  \gamma^{0},\,\gamma^{1},\,\gamma^{2},\,\gamma
^{3}\mathbf{,\,}\gamma^{5}\right)  \text{ ,}\\
\gamma^{0} &  =\left(
\begin{array}
[c]{cc}%
0 & 1\\
1 & 0
\end{array}
\right)  ,\;\mathbf{\vec{\gamma}=}\left(
\begin{array}
[c]{cc}%
0 & i\vec{\sigma}\\
-i\vec{\sigma} & 0
\end{array}
\right)  ,\;\gamma^{5}=\left(
\begin{array}
[c]{cc}%
1 & 0\\
0 & -1
\end{array}
\right)  \text{ ,}\nonumber\\
\mathbf{\vec{\gamma}} &  =\left(  \gamma^{1},\,\gamma^{2},\,\gamma^{3}\right)
\text{ ,}\nonumber
\end{align}
which satisfy
\begin{equation}
\left\{  \gamma^{M},\gamma^{N}\,\right\}  =2\delta^{MN}\text{ ,}%
\end{equation}
then our ``quantum constraint'' is
\begin{equation}
\left(  x_{M}\gamma^{M}-a\right)  \Phi\left(  x\right)  =0\text{
.}\label{quantum sphere4}%
\end{equation}
This implies
\begin{align}
\left(  x_{N}\gamma^{N}+a\right)  \left(  x_{M}\gamma^{M}-a\right)
\Phi\left(  x\right)   &  =0\text{ ,}\\
\left(  x_{M}x^{M}-a^{2}\right)  \Phi\left(  x\right)   &  =0\text{
,}\nonumber
\end{align}
thus the constraint
\begin{equation}
x_{M}x^{M}-a^{2}=0
\end{equation}
follows from the condition (\ref{quantum sphere4}). Explicitly, our ``quantum
constraint'' is
\begin{equation}
\left(
\begin{array}
[c]{cc}%
x_{5}-a & x_{0}+i\vec{x}\cdot\vec{\sigma}\\
x_{0}-i\vec{x}\cdot\vec{\sigma} & -x_{5}-a
\end{array}
\right)  \Phi\equiv\Delta\Phi=0\text{ .}%
\end{equation}

It should be noticed that the rank of the $4\times4$ matrix $\Delta$ is $2$
and the degree of the freedom of $\Phi$ is $2$. The general solution of the
constraint equation is
\begin{equation}
\Phi=v\phi\text{ ,}%
\end{equation}
where
\begin{equation}
v=\frac{1}{\sqrt{2a\left(  a+x_{5}\right)  }}\left(
\begin{array}
[c]{c}%
a+x_{5}\\
x_{0}-i\vec{x}\cdot\vec{\sigma}%
\end{array}
\right)  \text{ ,}%
\end{equation}
and
\begin{equation}
v^{\dag}v=1_{\left[  2\times2\right]  }\text{ .}%
\end{equation}
Using the variables
\begin{align}
z &  \equiv x_{0}-i\vec{x}\cdot\vec{\sigma}\text{ ,}\\
z^{\dag} &  \equiv x_{0}+i\vec{x}\cdot\vec{\sigma}\text{ ,}\nonumber
\end{align}
the induced gauge field is expressed as
\begin{equation}
A\equiv v^{\dag}dv=\frac{1}{2}\frac{1}{2a\left(  a+x_{5}\right)  }\left(
z^{\dag}dz-dz^{\dag}z\right)  \text{ ,}%
\end{equation}
which is the instanton gauge connection. In more familiar terms as
\begin{align}
z &  \equiv\alpha_{\mu}x_{\mu}\text{ ,}\\
\alpha_{\mu} &  \equiv\left(  1,-i\vec{\sigma}\right)  \text{ ,}\nonumber
\end{align}%
\begin{align}
z^{\dag} &  \equiv\bar{\alpha}_{\mu}x_{\mu}\text{ ,}\\
\bar{\alpha}_{\mu} &  =\left(  1,i\vec{\sigma}\right)  \text{ ,}\nonumber
\end{align}
we have
\begin{align}
\left(  z^{\dag}dz-dz^{\dag}z\right)   &  =\bar{\alpha}_{\mu}x_{\mu}%
\alpha_{\nu}dx_{\nu}-\bar{\alpha}_{\nu}dx_{\nu}\alpha_{\mu}x_{\mu
}\label{projection4}\\
&  =\left(  \bar{\alpha}_{\mu}\alpha_{\nu}-\bar{\alpha}_{\nu}\alpha_{\mu
}\right)  x_{\mu}dx_{\nu}\nonumber\\
&  \equiv2i\sigma_{\mu\nu}x_{\mu}dx_{\nu}\text{ ,}\nonumber
\end{align}
where
\begin{equation}
\frac{1}{2}\varepsilon_{\mu\nu\alpha\beta}\sigma_{\alpha\beta}=-\sigma_{\mu
\nu}.
\end{equation}
We finally arrive at
\begin{equation}
A=i\frac{1}{2a\left(  a+x_{5}\right)  }\sigma_{\mu\nu}x_{\mu}dx_{\nu}%
\end{equation}
which is the instanton gauge connection discussed in \cite{McMullan:1994ip}
\cite{Fujii:1995wn}. The projection operator in this case is
\begin{align}
P &  =vv^{\dag}\\
&  =\frac{1}{2a}\left(
\begin{array}
[c]{cc}%
a+x_{5} & x_{0}+i\vec{x}\cdot\vec{\sigma}\\
x_{0}-i\vec{x}\cdot\vec{\sigma} & a-x_{5}%
\end{array}
\right)  \text{ ,}\nonumber
\end{align}
and we have of course
\begin{equation}
P\partial\Phi=vD\phi\text{ .}%
\end{equation}

\section{Summary and discussion}

We have seen in this paper that quantum mechanics on $S^{2}$ and on $S^{4}$
can be reformulated by using the ``quantum constraint'', which is, in a sense,
a square root of the usual constraint $\vec{x}^{2}=a^{2}$ . Emergence of the
induced gauge fields (monopoles in $R^{3}$ or two dimensional $CP^{1}$
instantons on $S^{2}$, Yang monopoles in $R^{5}$ or instantons on $S^{4}$ )
can be attributed to the structure of projection operator.

We can read off from the wave function
\begin{equation}
\Phi=v\phi\text{ ,}%
\end{equation}
which can be rewritten as
\begin{equation}
Z=v\varphi:(Z^{\dagger}Z=\varphi^{\dagger}\varphi=1)\text{ ,}%
\end{equation}
where
\[
Z\equiv\frac{\Phi}{\sqrt{\Phi^{\dagger}\Phi}},\text{ }\varphi\equiv\frac{\phi
}{\sqrt{\phi^{\dagger}\phi}}\text{ ,}%
\]
the structure of the $S^{3}\rightarrow S^{2}$ $\left(  S^{7}\rightarrow
S^{4}\right)  $ Hopf fibering\cite{Minami:1979wn}\cite{Minami:1980pi}, where
$Z$ stands for the fiber bundle $S^{3}$\thinspace$\left(  S^{7}\right)  $, $v$
denotes the section $S^{2}\,\left(  S^{4}\right)  $ and finally $\varphi$ is
the fiber $U(1)\,\left(  SU(2)\right)  $.

Quantum mechanics on the odd dimensional spheres can be considered as a
reduction of the higher dimensional spheres, for example the $S^{3}$ case can
be obtained from the $S^{4}$ case by simply setting $x_{5}=0$ and we are left
with the induced meron gauge field\cite{Ikemori:1997fh}.

Finally a few comments as to the relation of our approach with that of the
coset space $G/H$ are in order. There is not so much difference as far as the
quantum mechanics on $S^{2}$ is concerned, however, the views seem rather
different on $S^{4}$. The structure of the fiber bundle looks different in
these two approaches, in our case it is not a group, has a simpler structure
and can be considered as a reduction of the former.

It is also interesting to note that our projection operator $P$
(\ref{projection4}) in the case of $S^{4}$ is related to that of ADHM
\cite{Corrigan:1984sv}\cite{Osborn:1982rd}\cite{Corrigan:1978ce}%
\cite{Dubois-Violette:1979it} construction of the instanton solutions to the
Yang-Mills gauge theory in its simplest case. The $S^{2}$ projection operator
on the other hand and particularly $v$ (\ref{v2}) plays a crucial role in
finding the instanton solution to the $CP^{1}$-$\sigma$ model.


\end{document}